\documentclass[twocolumn,twoside,preprintnumbers,amsmath,amssymb,showkeys]{revtex4}

\usepackage{graphicx}

\usepackage{fancyhdr}
\usepackage{pslatex}

\begin{document}
\title {Rapidity Dependence of HBT radii based on a hydrodynamical model}

\author{Kenji Morita$^1$\footnote{Present Address: Institute of Physics
and Applied Physics, Yonsei University, Seoul 120-749,
Korea. e-mail~:~\texttt{morita@phya.yonsei.ac.kr}}}

\affiliation{$^1$ Department of Physics, Waseda University, Tokyo
169-8555, Japan}

\begin{abstract}
 We calculate two-pion correlation functions at finite
 rapidities based on a hydrodynamical model which does
 not assume explicit boost invariance along the collision
 axis. Extracting the HBT radii through $\chi^2$ fits in
 both Cartesian and Yano-Koonin-Podgoretski\u{\i}
 parametrizations, we compare them with the experimental
 results obtained by the PHOBOS. Based on the results, we discuss
 longitudinal expansion	dynamics.
\end{abstract}

\vskip -1.35cm

\maketitle

\thispagestyle{fancy}

\setcounter{page}{1}

\bigskip

\section{INTRODUCTION}

``Perfect fluidity'' of the created matter at the Relativistic Heavy Ion
Collider (RHIC) in BNL is one of the most exciting news in the field of
high energy nuclear physics \cite{Perfectfluid}. Experimental results
and their comparison with theoretical calculation reveal that the
matter created at Au+Au collisions should be something like a liquid of
quarks and gluons, unlike a gas of almost free partons as naively
expected \cite{Whitepapers}. One of the strong evidences of this finding 
is an observation of large elliptic flow ($v_2$) and its agreement with
a perfect fluid-dynamical calculation \cite{Kolb_PLB500}. In order to
reproduce the experimental result with such models, an equation of state
assuming partonic state at high temperature and a phase transition and rapid
thermalization time ($\tau_0 \leq 1 \text{fm}/c$) are required
\cite{Kolb_PLB500}. 
Now the hydrodynamic model based on numerical solutions of the relativistic
hydrodynamic equation for perfect fluid becomes an indispensable tool for
theoretical analyses of the relativistic heavy ion collisions. 
Furthermore, the model itself has been becoming more sophisticated to
reproduce new experimental data with high statistics. Current most
sophisticated ones are full three-dimensional (solving hydrodynamic
equation without any symmetry) hydrodynamic expansion followed by a
hadronic cascade \cite{Hirano_PLB636,Nonaka_3dhybrid}. These models can
reproduce most of soft hadronic observables. Especially, simultaneous
description of particle ratios, transverse momentum spectra and elliptic
flow is possible with such hybrid models. 

However, there are still some insufficient ingredients in the
hydrodynamic analyses.
First, we don't have a reasonable initial condition derived from the
first principle. Recently, the Color Glass Condensate (CGC) has been
proposed as a suitable initial condition for relativistic heavy ion
collisions \cite{Iancu_QGP3}. This picture has been examined as an
initial condition for a hydrodynamic model in Ref.~\cite{Hirano_NPA743}
and found to give a good description of some observables in the case of
fully hydrodynamic description of the collision process. However, this
initial condition fails if one takes hadronic dissipation into account
\cite{Hirano_PLB636}. This fact suggests there is still an open space for
dissipative partonic phase, or improvement of the initial condition.

Second, the equation of state (EoS) of the QCD matter has not been fully
understood yet. Since one of the most important merit of using
hydrodynamic models is that it can be directly related to the EoS, detailed
information on the EoS for all region of temperature and baryonic chemical
potential is indispensable. As for the RHIC energy, net baryon number
observed at midrapidity is small enough to neglect it
\cite{BRAHMS_Netproton}. Nevertheless, EoS at finite baryonic chemical
potential may play important role at forward rapidity region and heavy
ion collisions at lower energies. Because of a well-known difficulty of
lattice QCD at finite baryonic chemical potential
\cite{Muroya_latticeQCD}, lattice QCD calculation has not provided the
complete solution yet. For vanishing baryonic chemical potential, the
lattice equation of state clearly shows a different behavior from the
free parton gas \cite{Aoki_LatticeEoS} and a lattice-inspired EoS is
implemented in a hydrodynamic calculation \cite{Hama_sph}.

At last, in spite of the success in most of soft observables, results of the
two-pion momentum intensity correlation from such hydrodynamical models
do not agree with
experimental data yet. According to the symmetry of the wave function of
two identical bosons, the two-particle correlation function can be
related to sizes of the source from which particles are emitted.
This fact is known as Hanbury Brown-Twiss (HBT) effect. Because it
concerns with source sizes, which depend on momentum of particle pairs
due to collective flow, the pion correlation function should be a
diagnostic tool for the space-time evolution of the matter. 
Since the disagreement was firstly found with a (2+1)-dimensional model
with boost
invariance along the collision axis \cite{Heinz_NPA702}, many extensions
such as an explicit longitudinal expansion
\cite{Morita_PRC66,Hirano_PRC66}, incorporating chemical freeze-out
\cite{Hirano_PRC66}, chiral model EoS \cite{Zschiesche_PRC65}, opaque
source \cite{Morita_PTP111}, fluctuating initial conditions and
continuous freeze-out \cite{Socorowski_PRL93} have been examined and the
discrepancy has been getting improved, but the situation is still
unsatisfactory. There are various possibilities for further improvements.

So far discussion on the HBT radii at the RHIC has been limited to
midrapidity because of acceptances of two experiment, STAR and PHENIX.
PHOBOS also has measured the two-pion correlation function.
By virtue of the wider acceptance of the detector, measurements at
non-zero rapidity windows can be done, and the data are now available in
Ref.~\cite{PHOBOS_HBT}. For analyses of such data in terms of the
Cartesian parameterization \cite{Bertsch_PRC37,Pratt_PRC42}, it should
be noted that there exists an additional HBT radius called ``out-long
cross term'' \cite{Chapman_PRL74} which vanishes at midrapidity due to
the symmetry. This radius contains information on the correlation
between  freeze-out points on the transverse plane and those on the
longitudinal direction. Hence, it is expected that this quantity is
sensitive to longitudinal expansion dynamics beyond boost-invariant
approximation. Similar consideration also holds for the
Yano-Koonin-Podgoretski\u{\i} parametrization which has three radius
parameters and one velocity parameter called YK velocity
\cite{YKP,Wu_EPJC1}. The PHOBOS data also provide rapidity dependence
of the YKP radii and YK velocity \cite{PHOBOS_HBT}, which may impose
a restriction on the longitudinal expansion dynamics. Indeed, the initial
matter distribution as an input for hydrodynamic calculations has not
been fixed yet. This is indicated by Hirano in
Ref.~\cite{Hirano_PRC65}, in which two different initial energy density
distributions can provide reasonable agreement with experimental data of
pseudorapidity distribution of charged hadrons measured in 130$A$ GeV
Au+Au collisions at RHIC.

In this work, we employ two different initial energy density
distributions for the hydrodynamic equations, as in
Ref.~\cite{Hirano_PRC65}. We focus our discussion on central
collisions. Both of them are so tuned that they reproduce
the pseudorapidity distribution of charged hadrons measured in the most
central events at 200$A$ GeV Au+Au collisions. Then we compare the
space-time evolution and shape of the freeze-out hypersurface of the
fluids and see how the difference in the initial condition is reflected
onto them. We calculate the two-pion correlation function as the most
promising experimental observable to see the difference. Extracting the
HBT radii through Gaussian fits, we compare them with the experimental
results and discuss the transverse momentum and rapidity dependence of
the HBT radii. In the next section, we briefly review the hydrodynamical
model used in this work. Initial conditions are given in
Sec.\ref{sec:initial}. In Sec.~\ref{sec:evolution}, we show numerical
solutions of hydrodynamical equations for the initial conditions given
in Sec.~\ref{sec:initial}. Results for the HBT radii as compared with the
experimental data are given in Sec.~\ref{HBT}. Section \ref{sec:summary}
is devoted to a summary.

\section{HYDRODYNAMICAL MODEL}\label{sec:model}

The basic equation of hydrodynamical models is the energy-momentum
conservation law
\begin{equation}
 \partial_\mu T^{\mu\nu}=0 \label{eq:hydro},
\end{equation}
where $T^{\mu\nu}$ is the energy-momentum tensor. For a perfect fluid,
\begin{equation}
 T^{\mu\nu}=(\varepsilon+P)u^\mu u^\nu - Pg^{\mu\nu} \label{eq:emtensor},
\end{equation}
where $g^{\mu\nu}=\text{diag}(+,-,-,-)$ and $\varepsilon$, $P$ and
$u^\mu$ are energy density, pressure and the four velocities of the fluid,
respectively. If one takes a conserved charge $i$ such as baryon number
and strangeness into account, the
conservation law
\begin{equation}
 \partial_\mu (n_i u^\mu) = 0 \label{eq:charge}
\end{equation}
is added to be solved. Providing an EoS $P=P(\varepsilon,n_i)$, one can
solve these coupled equations numerically.

In this work, we consider the baryon number charge as a conserved charge
and adopt an equation of state which exhibits a first order phase
transition on the phase boundary in $T-\mu_{\text{B}}$ plane from the
free massless partonic gas with three flavors to the free resonance gas
which consists of hadrons except for hyperons up to 2 GeV/$c^2$ of mass with
excluded volume correction\footnote{Note that, however, this EoS does
not agree with recent lattice QCD calculations which exhibit cross-over
transition at vanishing and small $\mu_{\text{B}}$
\cite{Aoki_LatticeEoS}. }. See Ref.\cite{Nonaka_EPJC17} for the
detail. The critical
temperature $T_{\text{c}}$ at vanishing baryonic chemical potential is
set to 160 MeV. This model of current use is basically same with the one
used in Refs.~\cite{Morita_PRC66,Morita_PTP111}.

Putting $z$-axis as the collision axis, we use a
cylindrical coordinate system $(\tau,\eta_{\text{s}},r,\phi)$ as follows;
\begin{align}
 t &= \tau\cosh\eta_{\text{s}}, \\
 z &= \tau\sinh\eta_{\text{s}}, \\
 r_x &= r \cos\phi, \\
 r_y &= r \sin\phi.
\end{align}
Here, $\tau=\sqrt{t^2-z^2}$ is the proper time and $\eta_{\text{s}}=1/2
\ln[(t+z)/(t-z)]$ is the space-time rapidity. Since we focus on central
collisions, we assume azimuthally symmetric system. Then, by virtue of
$u_\mu u^\mu=1$, the four velocities are given in terms of a
longitudinal flow rapidity $Y_{\text{L}}$ and a transverse flow rapidity
$Y_{\text{T}}$ as
\begin{align}
 u^{\tau} &= \cosh(Y_{\text{L}}-\eta_{\text{s}})\cosh Y_{\text{T}},
 \label{eq:utau}\\
 u^{\eta_{\text{s}}} &= \sinh(Y_{\text{L}}-\eta_{\text{s}})\cosh
 Y_{\text{T}}, \label{eq:ueta}\\
 u^r &= \sinh Y_{\text{L}}. \label{eq:ur}
\end{align}
To solve the equations numerically, we employed a method based on
the Lagrangian hydrodynamics which traces flux of the current.
The numerical procedure is described in Ref.~\cite{Ishii_PRD46}. 
For a treatment of the first order phase transition, we introduce a
fraction of the volume of the QGP phase to express the energy density
and net baryon number density at the mixed phase \cite{Nonaka_EPJC17}.
In this algorithm, we explicitly solve the entropy and baryon number
conservation law. We checked that these quantities are conserved
throughout the numerical calculation within 1\% of accuracy for a time
step $\delta\tau=0.01$ fm/$c$, by choosing proper mesh sizes of
$\eta_{\text{s}}$ and $r$ directions.

\section{INITIAL CONDITIONS}\label{sec:initial}

Firstly, we choose an initial proper time as $\tau_0=1$ fm/$c$. 
Initial values for other variables are given on this hyperbola.
Longitudinal flow rapidity is set to the
Bjorken's scaling ansatz $Y_{\text{L}}=\eta_{\text{s}}$
\cite{Bjorken_PRD27}. Transverse flow is simply neglected at the initial
proper time \footnote{Note that the existence of the initial transverse
flow can improve results for transverse momentum spectra but is an open
issue. In this work, however, we simply neglect it in order to avoid to
add additional parameters such as flow strength and profile. Since we
are mainly focusing on longitudinal expansion dynamics in this paper,
this simplification will not affect our main argument.}. For the
matter distributions, we
assume that the energy and
baryon number density are proportional to the number of binary
collisions. Hence, for the Woods-Saxon profile of nucleon density in
nuclei
\begin{equation}
 \rho_{\text{W}}(r,z)=\frac{\rho_0}{e^{(\sqrt{r^2+z^2}-R)/\xi}+1},
\end{equation}
where $R=1.12A^{1/3}-0.86A^{-1/3}$ fm is the radius of the nuclear with
mass number $A$, $\xi=0.54$ fm is the surface diffuseness and $\rho_0$
is the normal nuclear matter density, the density of binary collisions at
vanishing impact parameter is given by
\begin{equation}
 n_{\text{BC}}(r)=\sigma_0 
  \left[\int_{-\infty}^{\infty}dz
   \,\rho_{\text{W}}(r,z)\right]^2,\label{eq:bin}
\end{equation}
with $\sigma_0$ being the total inelastic nucleon-nucleon cross section
which is absorbed into the proportionality constant between
$n_{\text{BC}}$ and matter distributions.

Then, the energy density distribution is parameterized with a
``flat+Gaussian'' form, 
\begin{equation}
 \varepsilon(\tau_0, \eta_{\text{s}}, r) = \epsilon_0 
  \exp\left[-\frac{(|\eta_{\text{s}}|-\eta_{\text{s0}})^2}
       {2\sigma_{\eta_{\text{s}}}^2}
       \theta(|\eta_{\text{s}}|-\eta_{\text{s0}})\right]
  \overline{n_{\text{BC}}}(r). \label{eq:edist}
\end{equation}
Here, $\overline{n_{\text{BC}}}(r)$ is the normalized density of binary
collisions \eqref{eq:bin}, $\varepsilon_0$ the maximum energy density, and
$\eta_{\text{s0}}$ and $\sigma_{\eta_{\text{s}}}$ are parameters which
determine the length of the flat region and width of the Gaussian part,
respectively. Similarly, the net baryon number density distribution is
parameterized as
\begin{gather}
 n_{\text{B}}(\tau_0,\eta_{\text{s}},r)=n_{\text{B0}}
  \left\{
   \exp\left[
	-\frac{(|\eta_{\text{s}}|-\eta_{\text{sD}})^2}{2\sigma_{\text{sD}}^2}
      \right]\theta(|\eta_{\text{s}}|-\eta_{\text{s0}})\right. \nonumber \\
   \left. +\exp\left[
       -\frac{(\eta_{\text{s0}}-\eta_{\text{sD}})^2}{2\sigma_{\text{sD}}^2}
       \right]\theta(\eta_{\text{s0}}-|\eta_{\text{s}}|)\right\} 
  \overline{n_{\text{BC}}}(r), \label{eq:nbdist}
\end{gather}
where $n_{\text{B0}}$ is the maximum net baryon number density and
$\eta_{\text{sD}}$ and $\sigma_{\text{sD}}$ are the shape parameters as
in Eq.~\eqref{eq:edist}.

To calculate final particle distribution, we use the Cooper-Frye
prescription \cite{Cooper_PRD10}. The pseudorapidity distribution for
a particle species $i$ is given by
\begin{equation}
 \frac{dN_i}{d\eta} = \frac{d_i}{(2\pi)^2}\int_{0}^{\infty}dk_t 
  \frac{k_t |\mathbf{k}|}{k^0}\int_{\Sigma}k\cdot d\sigma 
  f(k\cdot u,T,\mu_{\text{B}}),\label{eq:dndeta}
\end{equation}
where $k^\mu$ is the momentum of thermally produced particles $i$ with
$d_i$ being the degree of freedom, pseudorapidity $\eta$ defined
by $\eta=1/2\ln[(|\mathbf{k}|+k_z)/(|\mathbf{k}|-k_z)]$ and 
$f(k\cdot u,T,\mu_{\text{B}})$ the equilibrium distribution
functions. We take into account not only directly
produced particles but also resonance decay contributions.
The freeze-out
hypersurface $\Sigma$ is chosen to be a constant temperature one,
$T=T_{\text{f}}=140$ MeV. 
Here, we assume thermal and chemical
freeze-out occur simultaneously. Since experimental data of particle
yields can be well described by the statistical model with high chemical
freeze-out temperature close to $T_{\text{c}}$
\cite{Braun-Munzinger_PLB518}, we cannot reproduce the correct particle
yields with this lower freeze-out temperature. However, in hydrodynamic
analyses, thermal freeze-out temperature is sensitive to $k_t$ spectra
which mainly reflects transverse expansion. In this calculation, we set
the freeze-out temperature so that pion $k_t$ spectrum is roughly
reproduced in IC B and set the same freeze-out temperature for IC A and
IC B. Note that the freeze-out temperature depends on the choice of the
transverse profile of
the initial matter distribution because a steeper pressure gradient yields
larger transverse flow. For example, even $T_{\text{f}} \simeq 150-160$ MeV is
possible with an initialization based on pQCD+saturation model
\cite{Eskola_PLB566}. Our value is only slightly different from
Ref.~\cite{Hirano_PRC65} where the initial profile is very similar.
In order to reproduce both of the particle yields and the
$k_t$ spectra in dynamical regimes, one should introduce separate
freeze-out temperatures \cite{Hirano_PRC66,Kolb_PRC67} or go to hybrid
approach \cite{Bass_PRC61,Teaney_PRL86,Hirano_PLB636,Nonaka_3dhybrid}.
In this work, however, our main argument will not be so affected by the
description of the freeze-out because we focus on longitudinal expansion.
\begin{table}[ht]
 \begin{center}
  \caption{Parameters in initial matter distributions}
  \label{tbl:initial}
  \begin{tabular}{ccccccc}\hline
   &$\varepsilon_0$
   [GeV/fm$^3$]&$\eta_{\text{s0}}$&$\sigma_{\eta\text{s}}$&$n_{\text{B0}}$
   [fm$^{-3}$]&$\eta_{s\text{D}}$&$\sigma_{s\text{D}}$ \\\hline
   IC.A&23.0&1.0&1.48&0.47&2.2&0.9 \\
   IC.B&20.5&3.0&0.33&0.55&2.2&0.75 \\ \hline
  \end{tabular}
 \end{center}
\end{table}

\begin{figure}[ht]
 \begin{center}
  \includegraphics[width=0.45\textwidth]{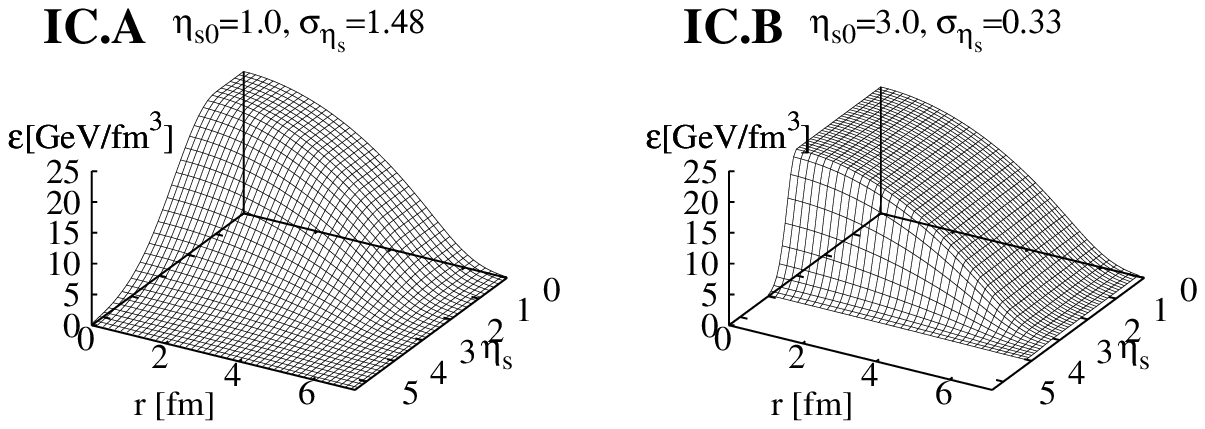}
  \caption{Initial energy density distributions}
  \label{fig:initiale}
  \includegraphics[width=0.45\textwidth]{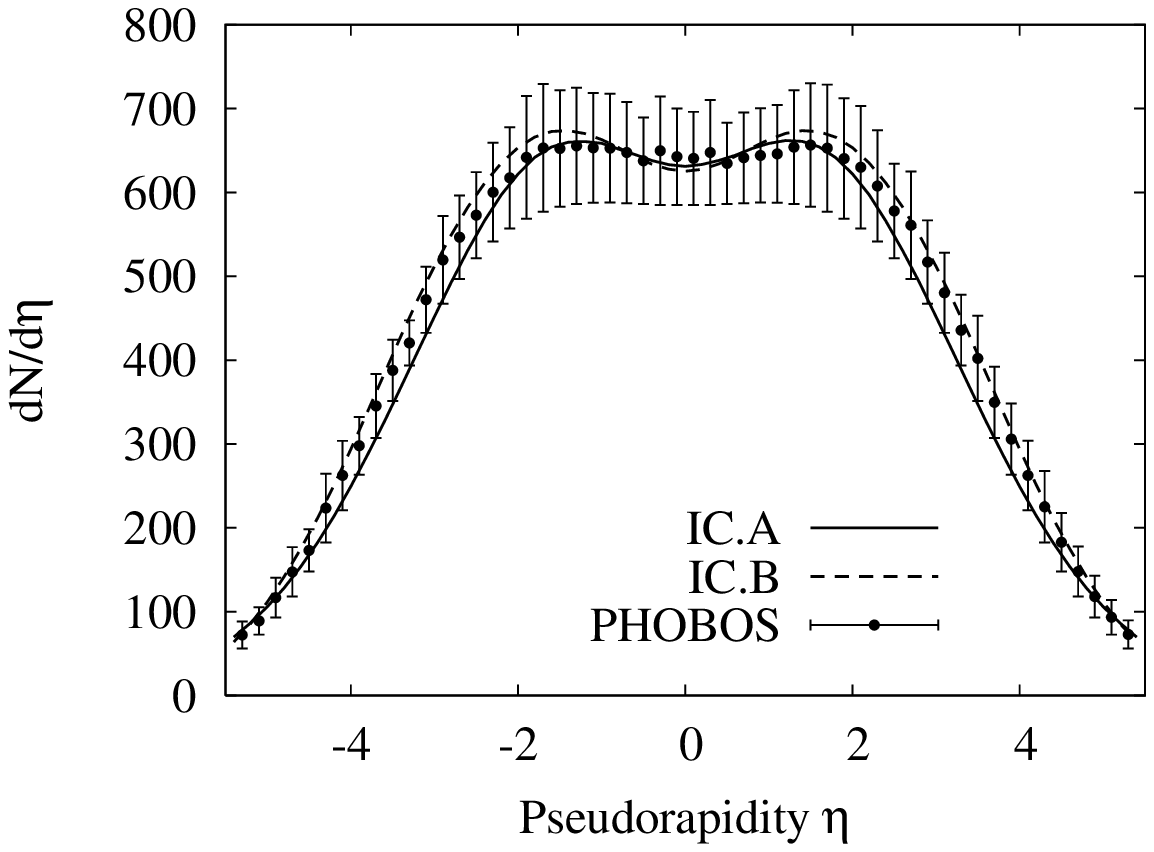}
  \caption{Pseudorapidity distribution. The solid line and dashed line
  stand for the initial condition A and B, respectively. Experimental
  data measured by PHOBOS are taken from Ref.\cite{PHOBOS_dndeta}}
  \label{fig:pseudorapidity}
  \includegraphics[width=0.45\textwidth]{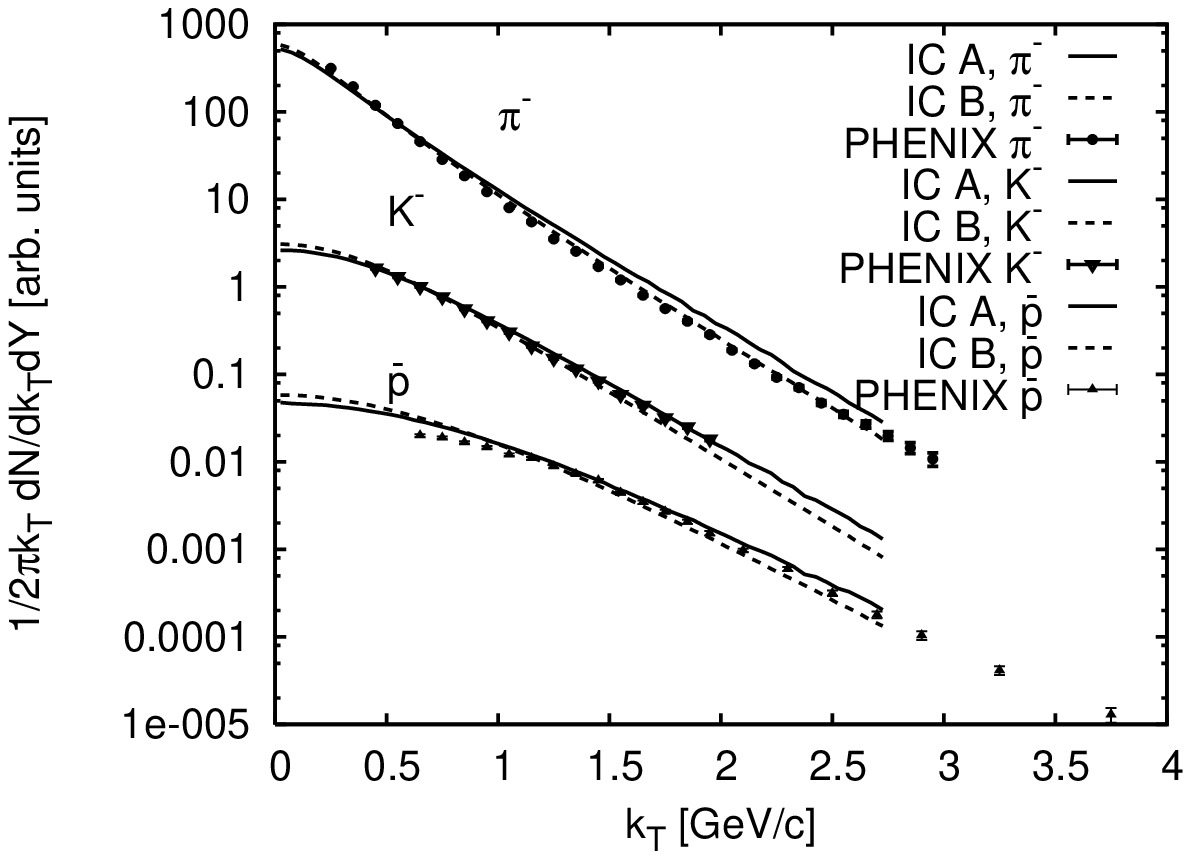}
  \caption{Transverse momentum distribution of identified negatively
  charged hadrons. Spectra for kaons and anti-protons are scaled by
  factor 0.1 and 0.01 for a clear comparison of the slopes,
  respectively. Experimental data measured by PHENIX are taken from
  Ref.~\cite{PHENIX_PRC69}. Identification of symbols is same as
  Fig.~\ref{fig:pseudorapidity}. }
  \label{fig:pt}
 \end{center}
\end{figure}
In Table \ref{tbl:initial}, two sets of the initial parameters are
listed. The corresponding initial energy density distributions,
resultant pseudorapidity distributions and transverse momentum
distributions are illustrated in
Fig.~\ref{fig:initiale},~\ref{fig:pseudorapidity} and \ref{fig:pt},
respectively. 
We have chosen two initial conditions both of which reproduce the
experimental data of PHOBOS on pseudo-rapidity distribution of charged
hadron \cite{PHOBOS_dndeta}, PHENIX on transverse momentum distributions
of $\pi^-$, $K^-$ and $\bar{p}$ \cite{PHENIX_PRC69}, and BRAHMS on
rapidity distribution of net protons \cite{BRAHMS_netproton}. 
These two initial conditions are characterized by two
parameters, $\eta_{\text{s0}}$ and $\sigma_{\text{s0}}$. One has small
$\eta_{\text{s0}}$ and large $\sigma_{\text{s0}}$, which we denote
initial condition A (IC.A). The other which we represent IC.B, has an
opposite feature; $\eta_{\text{s0}}$ is large and $\sigma_{\text{s0}}$
is small. The initial energy densities are both much larger than experimental
estimations ($\sim$ 5 GeV/fm$^3$) based on Bjorken's formula
\cite{Whitepapers}, but note that
$\varepsilon_0$ in Table \ref{tbl:initial} is not an \textit{average}
energy density but \textit{maximum} energy density, which strongly
depends on the profile of initial matter distributions
\cite{Kolb_NPA696}.
We calculated the pseudorapidity distributions for not only
these two initial conditions but also intermediate ones by varying
$\eta_{\text{s0}}$ from 1.0 to 3.0 and found that they can also
reproduce the experimental data by adjusting other parameters
appropriately. Perhaps the best fit will exist in the
middle of this parameter range \cite{Satarov_hep0606074}. Here, we
choose the extreme cases in order to see differences in the space-time
evolution of the fluids originating from the difference in the initial
conditions. 

\section{SPACE-TIME EVOLUTION OF THE FLUIDS}\label{sec:evolution}

\begin{figure}[ht]
 \begin{center}
  \includegraphics[width=0.45\textwidth]{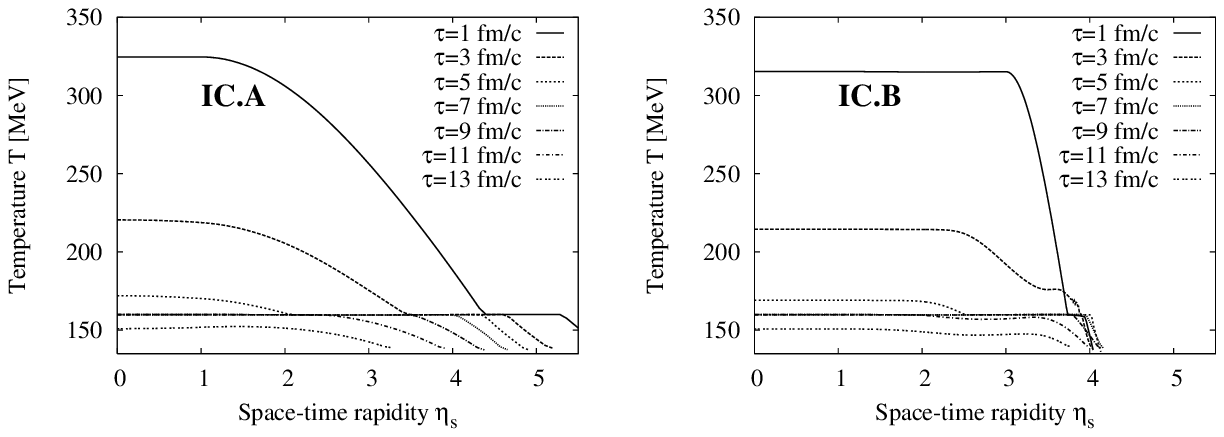}
  \caption{Space-time evolution of temperature distributions at $r=0$. }
  \label{fig:T-eta}
  \includegraphics[width=0.45\textwidth]{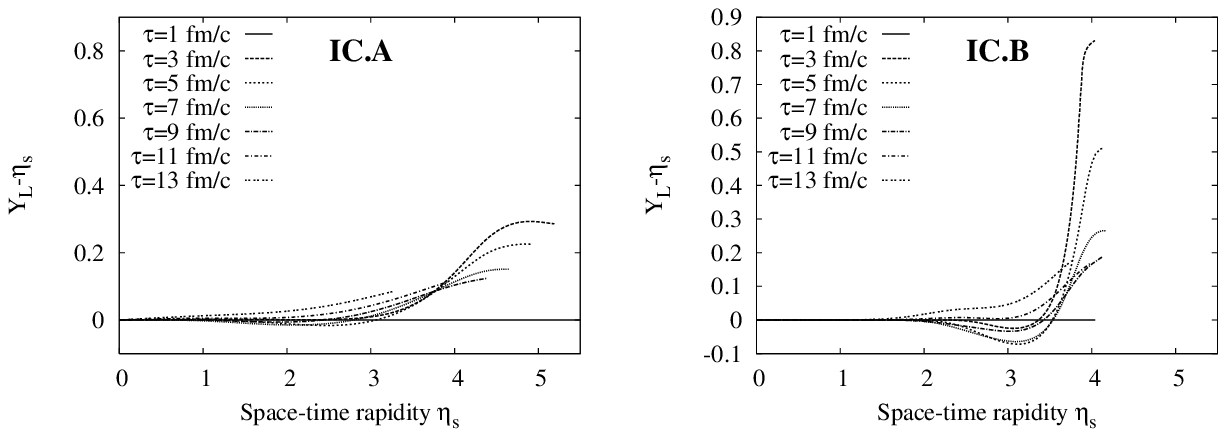}
  \caption{Deviation from the scaling solution at $r=0$.}
  \label{fig:yl-eta}
 \end{center}
\end{figure}

Figures \ref{fig:T-eta} and \ref{fig:yl-eta} show the space-time
evolution of the temperature distributions and  deviation from the
scaling solution $Y_{\text{L}}=\eta_{\text{s}}$, as a function of
$\eta_{\text{s}}$ at $r=0$ for various $\tau$, respectively.
From these figures, we find that the space-time evolution at forward
rapidity is quite different between IC.A and IC.B in spite of the fact
that both solutions give similar pseudorapidity distributions of
hadrons. In IC.B, sharp decrease of temperature which is identical to
steep pressure gradient at the forward rapidity causes rapid
acceleration of the longitudinal flow at the edge of the fluid. 
On the other hand, in IC.A, pressure gradient is rather gradual.
Hence, resultant deviation from the scaling solution is smaller.
Because pressure gradient exist at smaller $\eta_{\text{s}}$ in IC.A,
however, such deviations take place at $\eta_{\text{s}}\simeq 1$ while 
the flow keeps the scaling solutions up to $\eta_{\text{s}}\simeq 2$ in
IC.B. This fact explains slightly larger $\varepsilon_0$
in IC.A since faster longitudinal expansion than the scaling expansion pushes
entropy per unit rapidity to forward rapidity
\cite{Eskola_EPJC1,Morita_PRC66}.

Although Figs.~\ref{fig:T-eta} and \ref{fig:yl-eta} show that there
exist differences between IC.A and IC.B in the space-time evolution,
it is not trivial that such differences can survive at the freeze-out
hypersurfaces. Since hadrons strongly interact and have information only
at the thermal freeze-out, differences on the freeze-out hypersurfaces
are necessary to find the signature in hadronic experimental
observables. 

\begin{figure}[ht]
 \begin{center}
  \includegraphics[width=0.45\textwidth]{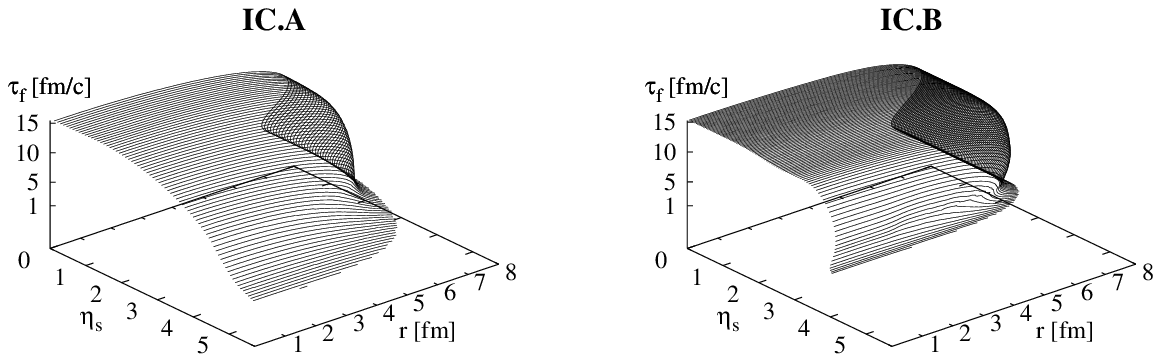}
  \caption{Freeze-out hypersurface of the fluids.}
  \label{fig:surface}
  \includegraphics[width=0.45\textwidth]{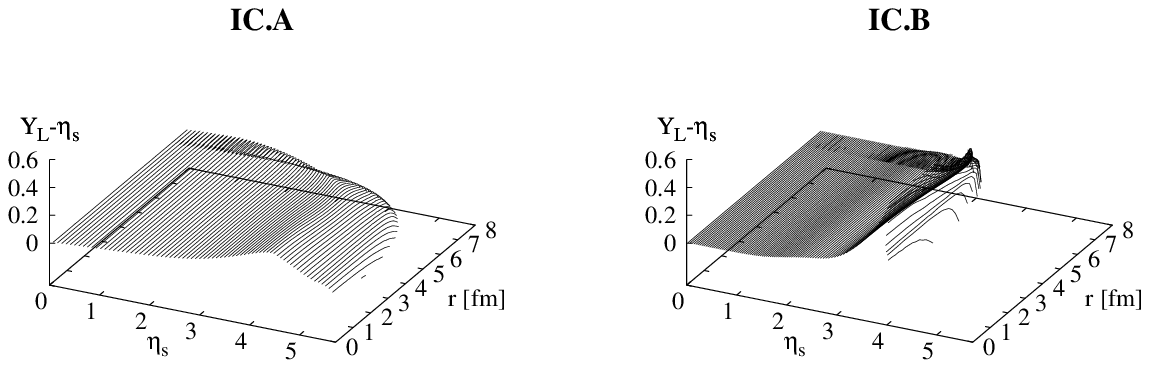}
  \caption{Deviation from the scaling solution on the freeze-out
  hypersurfaces.}
  \label{fig;yl-surface}
 \end{center}
\end{figure}

We show freeze-out proper time $\tau_{\text{f}}$ of the all fluid
elements in Fig.~\ref{fig:surface}. This characterizes the shape of the
freeze-out hypersurface, which is expected to affect the HBT radii.
In Fig.~\ref{fig:surface}, we can see that the system expands in the
transverse direction in both of the fluids. Due to the same transverse
profile, there is no apparent difference in the transverse direction.
On the other hand, the shape of the hypersurface in the
$\eta_{\text{s}}$ direction shows some variations. In IC.B, expansion
appears and the freeze-out proper time is mostly constant in the broad
range of $\eta_{\text{s}}$ while it moderately decreases with
$\eta_{\text{s}}$ in IC.A. This is a consequence of the different
longitudinal flow profile (Fig.~\ref{fig:yl-eta}). We also plot the
deviation from the scaling solution at the freeze-out in
Fig.~\ref{fig;yl-surface}. The large deviation seen at forward rapidity in
IC.B (Fig.~\ref{fig:yl-eta}) survives at the freeze-out. We will see how
these differences affect the HBT radii in the next section.

\section{HBT RADII}\label{HBT}
\subsection{Two-pion correlation function}

Assuming that the source is completely chaotic, we can calculate the
two-particle correlation momentum intensity correlation function through
this formula \cite{Shuryak_PLB44}
\begin{equation}
 C_2(q,K)=1+\frac{|I(q,K)|^2}{I(0,k_1)I(0,k_2)}, \label{eq:cor}
\end{equation}
where $q=k_1-k_2$ is the four-relative momentum and $K=1/2(k_1+k_2)$ is
the four-average momentum, with $k_i$ being on-shell momentum of emitted
pions. The interference term $I(q,K)$ can be chosen as
\begin{equation}
 I(q,K)=\int_\Sigma K\cdot d\sigma \, e^{iq\cdot x} f(u\cdot K,T),
  \label{eq:interf}
\end{equation}
so that $I(0,k_i)$ reduces to the Cooper-Frye formula
\cite{Chapman_PLB340}.

Experimentally, the two-pion correlation function is defined as
\begin{equation}
 C_2(\mathbf{q})=\frac{A(\mathbf{q})}{B(\mathbf{q})},
\end{equation}
where $A(\mathbf{q})$ is the measured two-pion pair distribution with
momentum difference $\mathbf{q}$ and $B(\mathbf{q})$ is the background
pair distribution generated from mixed events. Momentum acceptances are
imposed separately in the numerator and the denominator. Accounting for
the large acceptance in the PHOBOS experiment, $0.4 < Y_{\pi\pi} < 1.3$
for three $K_{\text{T}}$ bins and $0.1 < K_{\text{T}} < 1.4$ GeV/$c$ for
three rapidity bins, we integrate the correlation function as follows:
\begin{align}
 C(q;K_{\text{T}}) &= 1+\frac{\int_{0.4}^{1.3}dY_{\pi\pi}
 |I(q,K)|^2}{\int_{0.4}^{1.3} dY_{\pi\pi} I(0,k_1)I(0,k_2)}
 ,\label{eq:c2kdep}\\
 C(q;Y_{\pi\pi}) &= 1+\frac{\int_{0.1}^{1.4}dK_{\text{T}}\,K_{\text{T}}\,
 |I(q,K)|^2}{\int_{0.1}^{1.4} dK_{\text{T}}\,K_{\text{T}}\, I(0,k_1)I(0,k_2)}.\label{eq:c2ydep}
\end{align}
For simplicity, we consider only directly emitted pions and neglect
resonance decay contributions.

\subsection{$K_{\text{T}}$ dependence of the HBT radii in the Cartesian
  parametrization}

Physical meaning of the HBT radii depends on the choice of three independent
components of the relative momentum $q$. The most standard choice is the
so-called Cartesian Bertch-Pratt parametrization
\cite{Bertsch_PRC37,Pratt_PRC42}
$\mathbf{q}=(q_{\text{out}}, q_{\text{side}}, q_{\text{long}})$ in which
``long'' means parallel to the collision axis, ``side'' perpendicular to
the transverse component of the average momentum
$\mathbf{K_{\text{T}}}$ and ``out'' parallel to $\mathbf{K_{\text{T}}}$.
In the case of azimuthally symmetric system as considered here, one can
put $\mathbf{K_{\text{T}}}=(K_{\text{T}},0)$ so that
$q_{\text{out}}=q_x$ and $q_{\text{side}}=q_y$. Note that
$q_{\text{long}}=q_z$. Then, the Gaussian form of the two-pion
correlation function is given as \cite{Chapman_PRL74}
\begin{gather}
 C_{2\text{fit}}(\mathbf{q})=1+\lambda \exp(-q_{\text{out}}^2
  R_{\text{out}}^2-q_{\text{side}}^2 R_{\text{side}}^2 -
  q_{\text{long}}^2 R_{\text{long}}^2 \nonumber\\ -2
  q_{\text{out}}q_{\text{long}}R_{\text{ol}}^2).\label{eq:cartesianc2}
\end{gather}
The HBT radii $R_i$ can be extracted by a $\chi^2$-fit to the above
fitting function. For a chaotic source, the chaoticity parameter
$\lambda$ should become unity. However, experimentally observed
chaoticity is smaller than 1 because of such contributions as long-lived
resonance decay \cite{Morita_3pi}. Here we fix $\lambda=1$ in the
Gaussian fit to the calculated correlation functions with
Eqs.~\eqref{eq:c2kdep} and \eqref{eq:c2ydep}.

By expanding the correlation function \eqref{eq:cor} for $q\cdot x \ll
1$,  the size parameters $R_i$ can be related to second order
 moments of the source function \cite{Chapman_PRL74}. In the Cartesian
 parametrization, taking the longitudinal co-moving system (LCMS) makes
 the expression simple;
 \begin{align}
  R_{\text{out}}^2 &= \langle (\tilde{r_x}-\beta_{\perp}\tilde{t})^2
  \rangle \nonumber \\
  &= \langle \tilde{r_x}^2 \rangle -2\beta_{\perp}
  \langle \tilde{r_x}\tilde{t} \rangle 
  +\beta_{\perp}^2 \langle \tilde{t}^2 \rangle. \label{eq:rout} \\
  R_{\text{side}}^2 &= \langle \tilde{r_y}^2 \rangle, \\
  R_{\text{long}}^2 &= \langle \tilde{z}^2 \rangle,\\
  R_{\text{ol}}^2 &= \langle
  (\tilde{r_x}-\beta_{\perp}\tilde{t})\tilde{z} \rangle,
 \end{align}
where
\begin{equation}
 \langle A(x) \rangle \equiv 
  \frac{\int_\Sigma k\cdot d\sigma f(u\cdot k,T) A(x)}
  {\int_\Sigma k\cdot d\sigma f(u\cdot k,T)},
\end{equation}
$\tilde{x}\equiv x-\langle x \rangle$, and
$\beta_{\perp}=k_{\text{T}}/E_{\mathbf{k}}$. Hence, $R_{\text{out}}$,
$R_{\text{side}}$ and $R_{\text{long}}$ can be interpreted as a mixture
of thickness of the source and emission duration, transverse source
size and longitudinal source, seen from the LCMS, respectively.
Validity of these expressions for a hydrodynamical model is discussed in
Ref.~\cite{Morita_PRC61}. Although they have been shown to be good
approximations, it is also pointed out that there are still some
discrepancies and one should use fitted HBT radii for comparison with
the experimental data which are obtained from the fit
\cite{Frodermann_PRC73}.

\begin{figure}[ht]
 \begin{center}
  \includegraphics[width=0.34\textwidth]{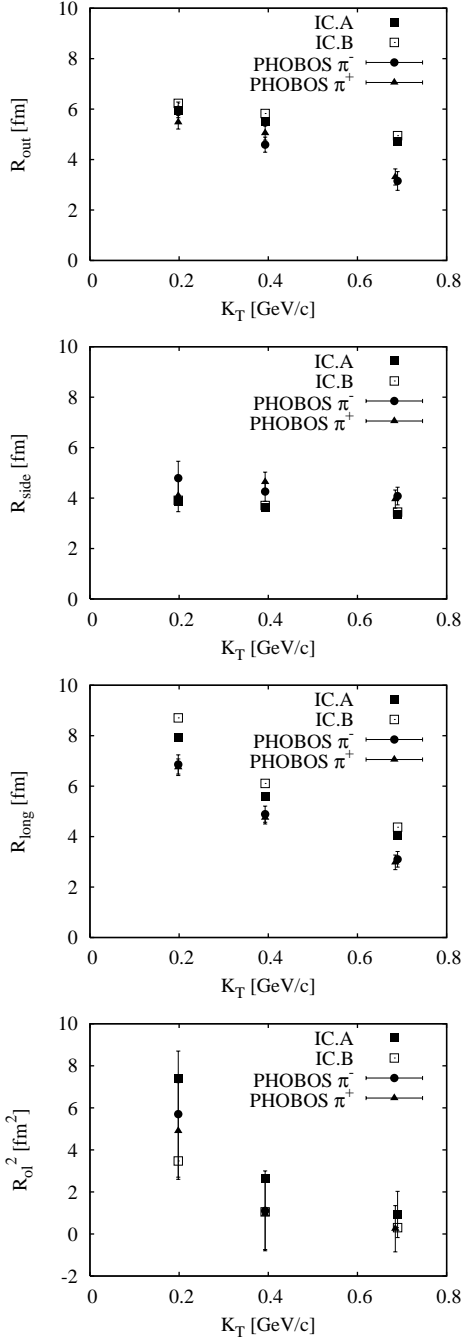}
  \caption{$K_{\text{T}}$ dependence of Cartesian HBT radii. Closed
  squares and open squares denote our results for IC.A and IC.B,
  respectively. Experimental data are taken from
  Ref.~\cite{PHOBOS_HBT}. Error-bars for the experimental data are
  statistical only.}
  \label{fig:hbt_cart}
 \end{center}
\end{figure}

Figure \ref{fig:hbt_cart} shows results for the four HBT radii
compared with the experimental data measured by PHOBOS
\cite{PHOBOS_HBT}. For comparison of the initial conditions, any
qualitative and quantitative difference cannot be seen in
$R_{\text{out}}$ and $R_{\text{side}}$, as expected from
Fig.~\ref{fig:surface}. $R_{\text{long}}$ of IC.A is about 1 fm smaller
than that of IC.B. This can be considered as a consequence of the fact
that the deviation from the scaling solution at small $\eta_{\text{s}}$
is larger in IC.A, because faster flow causes more thermal suppression
of the emission region \cite{Morita_PRC66}. For these three radii, our
calculation cannot reproduce the experimental results and show similar
behavior with other perfect fluid dynamical calculations of
Ref.\cite{Heinz_NPA702,Morita_PRC66,Morita_PTP111,Hirano_PRC66}.
Especially
$R_{\text{long}}$ shows the largest deviation from the experimental
data, although calculation is improved by the longitudinal
expansion without explicit boost invariance \cite{Morita_PRC66,Hirano_PRC66}.
In the bottom of Fig.~\ref{fig:hbt_cart}, result of the out-long cross
term is presented. Reflecting the uniform shape of the freeze-out
hypersurface in Fig.~\ref{fig:surface}, the value of $R_{\text{ol}}^2$
of IC.B is smaller than that of IC.A. At the lowest $K_{\text{T}}$ bin, 
the difference is about 4 fm$^2$. Unfortunately, experimental
uncertainty is still too large to distinguish which initial condition is
favored. However, it should be noted that both of two results agree with
the experimental data, in spite of the disagreement of other radii.

\subsection{Rapidity dependence of the HBT radii in the YKP
  parametrization}

In the YKP parametrization, three independent components of the relative
momentum $q$ are $q_{\perp}=\sqrt{q_x^2+q_y^2}$,
$q_\|=q_z=q_{\text{long}}$ and $q_\tau=E_1-E_2$. Then, the Gaussian fitting
correlation function is given as
\begin{gather}
 C_{\text{2YKP}}(\mathbf{q})=1+\lambda\exp\left[-R_{\perp}^2q_{\perp}^2
 -R_{\|}^2(q_{\|}^2-q_{\tau}^2)\right. \nonumber
 \\ \left. -(R_{\tau}^2+R_\|^2)(q\cdot U)^2\right],
 \label{eq:c2ykp}
\end{gather}
where $U^\mu = \gamma(1,0,0,v_{\text{YK}})$,
$\gamma=1/\sqrt{1-v_{\text{YK}}^2}$ and $v_{\text{YK}}$ is the fourth
fitting parameter called YK velocity. The three HBT radii, $R_{\perp}$,
$R_{\|}$ and $R_\tau$ are invariant under a longitudinal boost. Physical
meaning of the parameters can be given in a similar manner
\cite{Wu_EPJC1} and becomes the simplest as follows, if one adopt the YK
frame where $v_{\text{YK}}=0$,
\begin{align}
 R_{\perp}^2 &= \langle \tilde{r_y}^2 \rangle = R_{\text{side}}^2
 \label{eq:YKP-perp},\\
 R_{\|}^2 &\simeq \langle \tilde{z}^2 \rangle = R_{\text{long}}^2 
 \label{YKP-para},\\
 R_{\tau}^2 &\simeq \langle \tilde{t}^2 \rangle \label{YKP-tau}.
\end{align}
The main advantage of using YKP parametrization is that the three HBT
radii directly give the transverse, longitudinal and temporal source
size, that are seen from the YK frame. However, one should note that
the latter two, \eqref{YKP-para} and \eqref{YKP-tau}, are approximate
expressions which hold only if the source is not opaque
\cite{Morita_PRC61}. Hence, $R_{\|}$ and $R_{\tau}$ cannot be always
regarded as the source sizes in the presence of strong transverse flow
which makes the source highly opaque \cite{Morita_PTP111}. The general
expression of $v_{\text{YK}}$ is complicated one \cite{Wu_EPJC1} but it
can be regarded as a longitudinal flow velocity of the source measured
in an observer's frame.

\begin{figure}[ht]
 \begin{center}
  \includegraphics[width=0.35\textwidth]{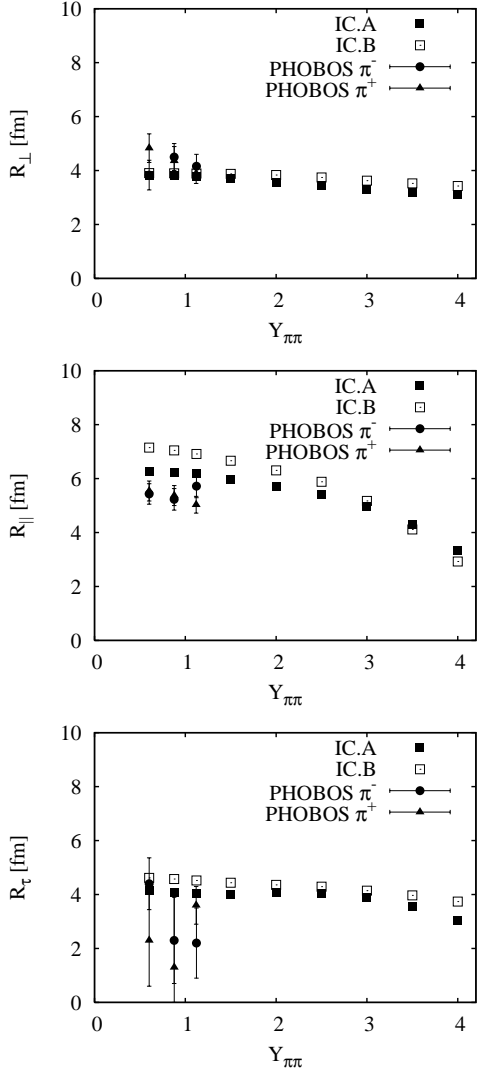}
  \caption{HBT radii for the YKP parametrization. The identification of
  the symbols is the same as in Fig.~\ref{fig:hbt_cart}.}
  \label{fig:ykp}
 \end{center}
\end{figure}

We plot results of HBT radii for the YKP parametrization in
Fig.~\ref{fig:ykp}. Though PHOBOS measures only at small values of 
rapidity, we calculate the HBT radii for
$Y_{\pi\pi}=0.602,0.877,1.122,1.5,2.0,2.5,3.0,3.5$ and $4.0$ and show
the results as a prediction. For comparison between IC.A and IC.B,
$R_{\perp}$ seems to barely reflect the uniform structure along
$\eta_{\text{s}}$ direction in IC.B. While $R_\|$ shows the difference
of order 1 fm at small rapidity coming from the deviation of the scaling
solution as well as in the third panel of Fig.~\ref{fig:hbt_cart}, $R_{\tau}$
shows little difference but agrees with the experiment. Large
experimental errors will be due to the known difficulty of the limited
kinematic region in the YKP parameterization \cite{Tomasik_ActaSlov}.
Because of the large $K_{\text{T}}$ window of the data, it is difficult
to estimate geometrical opacity effect on $R_\tau$. If we assume this
effect is small, a possible origin of the deviation of our result from
the data is larger emission duration. Some model calculations based on
source parametrization \cite{Retiere_PRC70} and parametric exact
solution of hydrodynamics \cite{Csanad_NPA742} show very
small emission duration time, 0-2 fm/$c$ in agreement with data on the
bottom panel of Fig.~\ref{fig:ykp}.
We cannot see any significant
differences in the HBT radii at forward rapidity expected from
Figs.~\ref{fig:surface} and \ref{fig;yl-surface} which display the
differences of the source shape and the longitudinal flow. This will
come from the fact
that the number of produced particles is larger at late freeze-out
proper time in the case of the current freeze-out condition
\cite{Morita_PRC61}.

\begin{figure}[ht]
 \begin{center}
  \includegraphics[width=0.45\textwidth]{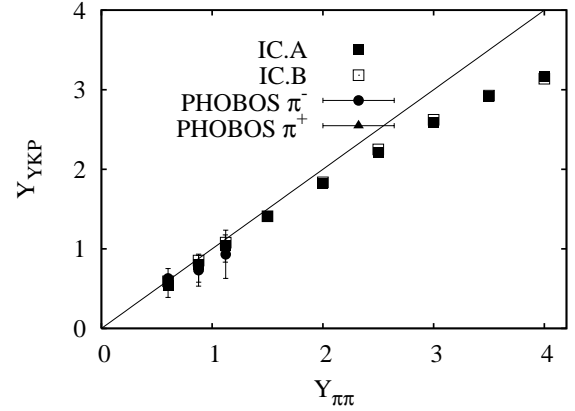}
  \caption{The Yano-Koonin rapidity $Y_{\text{YK}}$. The identification
  of the symbols is the same as in Figs.~\ref{fig:hbt_cart} and
  \ref{fig:ykp}. The solid line indicates the case of the infinite
  boost-invariant source.}
  \label{fig:Yykp}
 \end{center}
\end{figure}

Finally the Yano-Koonin rapidity
$Y_{\text{YK}}=1/2\ln[(1+v_{\text{YK}})/(1-v_{\text{YK}})]$ is shown as
a function of $Y_{\pi\pi}$ in Fig.~\ref{fig:Yykp}. Both of results from
IC.A and IC.B surprisingly agree with the experimental data and show no
difference between the two. At forward rapidity region, our results show
deviation from the infinite boost invariant case, which is indicated by
the straight line. Although our solutions of longitudinal flow show
deviation from the scaling solution (Figs.~\ref{fig:yl-eta} and
\ref{fig;yl-surface}), the result would have to larger $Y_{\text{YK}}$
than a given $Y_{\pi\pi}$ if $Y_{\text{YK}}$ correctly represents the
longitudinal source velocity. Hence, this deviation will be caused by
the finite size effect \cite{Morita_PRC61} which becomes more
significant at forward rapidity rather than the difference in the flow
velocity.

\section{SUMMARY}\label{sec:summary}

In summary, we calculated the two-pion correlation function for two
sources which are given by a hydrodynamical model without explicitly
boost invariance along the collision axis. The two initial conditions
are so chosen that both of them give consistent pseudorapidity
distribution with the experimental data and have different shape in the
longitudinal direction. Other model ingredient, initial transverse
profile from the binary collision model, scaling solution for initial
longitudinal flow, vanishing initial transverse flow, EoS with first order
phase transition and Cooper-Frye freeze-out prescription with
$T_{\text{f}}=140$ MeV are the same in the two solutions.
We find that there exist some differences in the
space-time evolution of the fluids in spite of the fact that both fluids
give similar particle distribution. The HBT radii are extracted from the
two-pion correlation functions and compared with the experiment.
In the Cartesian parametrization, the out-long cross term which arises
at nonzero rapidities shows a difference between two initial conditions
and the good agreements with the experimental data. The correlation
function is also analyzed with the YKP parametrization. We find a small
difference between the two initial conditions in $R_\|$ which reflects
deviation from the scaling solution in the longitudinal expansion as
well as $R_{\text{long}}$ in the Cartesian parametrization.
Possible sources of this disagreement are followings:
EoS of current use exhibits first order phase transtion which makes the
lifetime of the fluid longer, and assumes hadronic states is in fully
chemical equilibrium. It is known that both crossover EoS
\cite{Zschiesche_PRC65} and incorporating chemical freeze-out
\cite{Hirano_PRC66} improve the lifetime then
$R_{\text{long}}$ and $R_\|$. We used the conventional Cooper-Frye
prescription for the freeze-out. Improvement of the freeze-out
prescription by continuous freeze-out \cite{Hama_sph} and hybrid
approach \cite{Hirano_PLB636,Nonaka_3dhybrid} can yield larger
$R_{\text{side}}$ but this may lead to larger $R_{\text{out}}$ because
of extended emission duration. Nevertheless, as a transport calculation
\cite{Lin_PRL89} shows, positive $x-t$ correlation in the source
function may resolve this problem.
Finally, in spite of the disagreement of the HBT radii, the YK rapidity shows
a good agreement with the experimental data. Our calculation predicts 
some deviations at larger rapidities from the infinite boost-invariant case.
Hence, measurements at this region is needed for further understanding
of the expansion dynamics.

\bigskip
\noindent\textbf{Acknowledgements}
\medskip

The author is indebted to Profs. I.~Ohba and H.~Nakazato for their
encourgement. He also would like to thank S.~Muroya and T.~Hirano for
their helpful discussions. This work was supported by a Grant for the
21st Century COE Program at Waseda University from Ministry of
Education, Culture, Sports, Science and Technology of Japan.
This is also supported by BK21 (Brain Korea 21) program of the Korean
Ministry of Education.

\end{document}